\documentclass[twocolumn,showpacs,aps,pre,10pt]{revtex4-1}

\usepackage{balance}
\usepackage{lastpage}
\usepackage{flushend}
\usepackage{array}
\usepackage{xcolor}
\usepackage{graphicx}
\usepackage{amsmath}
\usepackage[abs]{overpic}
\usepackage{pgfplots}
\usepackage{flushend}
\newcolumntype{C}[1]{>{\centering\let\newline\\\arraybackslash\hspace{0pt}}m{#1}}

\begin{document}

\title{Distribution of Topological Types in Grain-Growth Microstructures}  

\author{Emanuel A. Lazar$^1$, Jeremy K. Mason$^2$, Robert D. MacPherson$^3$, David J. Srolovitz$^4$}
\affiliation
{$^1$Department of Mathematics, Bar-Ilan University, Ramat Gan 5290002, Israel\\
 $^2$Department of Materials Science and Engineering, University of California, Davis, California 95616, USA\\ 
 $^3$School of Mathematics, Institute for Advanced Study, Princeton, New Jersey 08540, USA\\
 $^4$Department of Materials Science and Engineering, City University of Hong Kong, Hong Kong SAR China
} 
\date{\today}

\begin{abstract}
An open question in studying normal grain growth concerns the asymptotic state to which microstructures converge.  In particular, the distribution of grain topologies is unknown.  We introduce a thermodynamic-like theory to explain these distributions in two- and three-dimensional systems.  In particular, a bending-like energy $E_i$ is associated to each grain topology $t_i$, and the probability of observing that particular topology is proportional to $\frac{1}{s(t_i)}e^{-\beta E_i}$, where $s(t_i)$ is the order of an associated symmetry group and $\beta$ is a thermodynamic-like constant.  We explain the physical origins of this approach, and provide numerical evidence in support.


\end{abstract}

\maketitle

{\it Introduction.} Theory, simulation, and experimental work have shown that during normal grain growth, polycrystalline microstructures evolve toward an asymptotic state in which scale-invariant properties become constant \cite{1965hillert, humphreys2012recrystallization}.  It has also been observed that this state is reached largely independently of initial conditions \cite{1986mullins, atkinson1988overview}.  A major goal in this field has therefore been to characterize and understand this universal grain-growth microstructure.  In addition to its geometric features \cite{1989anderson, kamachali2015geometrical, miyoshi2017ultra}, its topological features have also been carefully studied.  In two dimensions, grains can be classified by their number of edges \cite{1994fradkov, 1999fayad, meng2015study}.  An analogous approach is insufficient in three dimensions, as grains with the same number of faces can have distinct topologies.  Recent work has focused on characterizing the types of grain faces \cite{keller2014comparative}, the arrangements of those faces \cite{2012lazar, xue2016matrix, lutz2017roundness, perumal2017phase, wang2019finding}, and the manner in which edges are arranged in the grain boundary network \cite{2012mason}.

Previous studies characterizing the distribution of grain topologies is limited in two important ways.  First, little connection has been made between two- and three-dimensional systems; a general theory explaining both is desirable.  Second, despite careful characterization of grain types that appear and their relative frequencies, an explanation of these observations remains elusive.

This letter introduces a novel, thermodynamic-like approach to explain the observed distributions of topological types in two- and three-dimensional grain-growth systems.  In particular, we associate a bending-like energy to each grain that depends only on its topology, and show that this energy can help predict the distribution of topologies in these cellular microstructures.

\begin{figure}[b]
\begin{center}    
\begin{tabular}{cc}      
\resizebox{0.25\columnwidth}{!}{\includegraphics{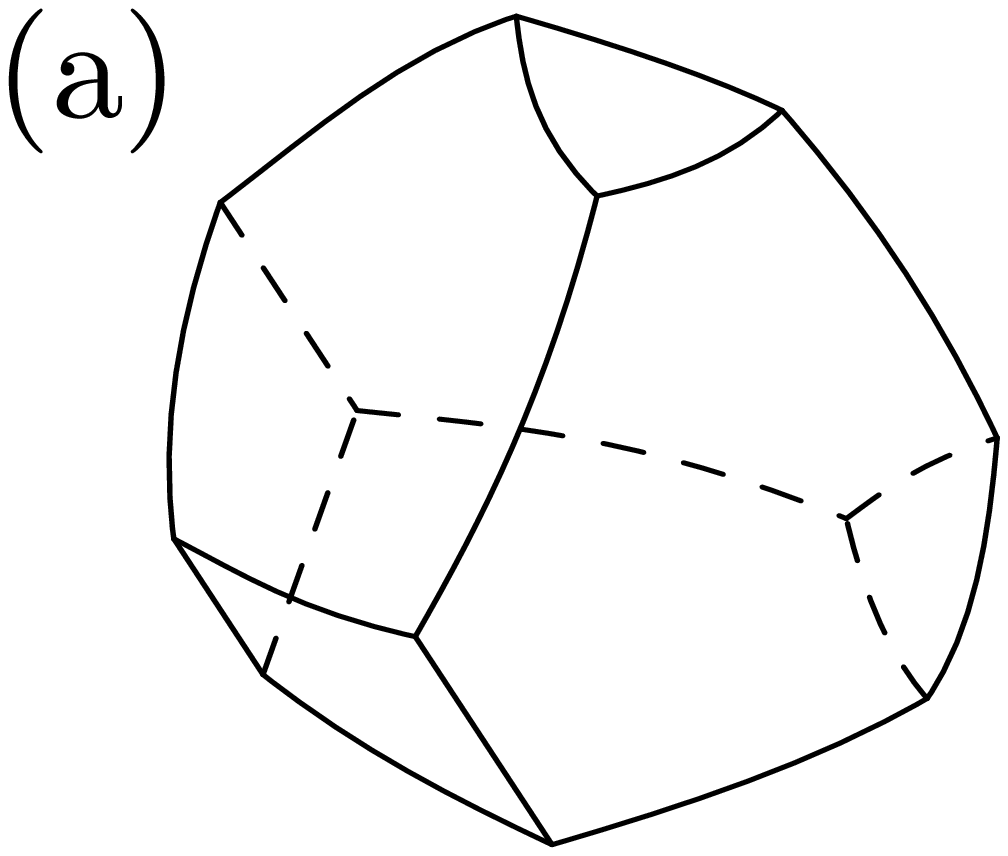}} &\quad
\resizebox{0.25\columnwidth}{!}{\includegraphics{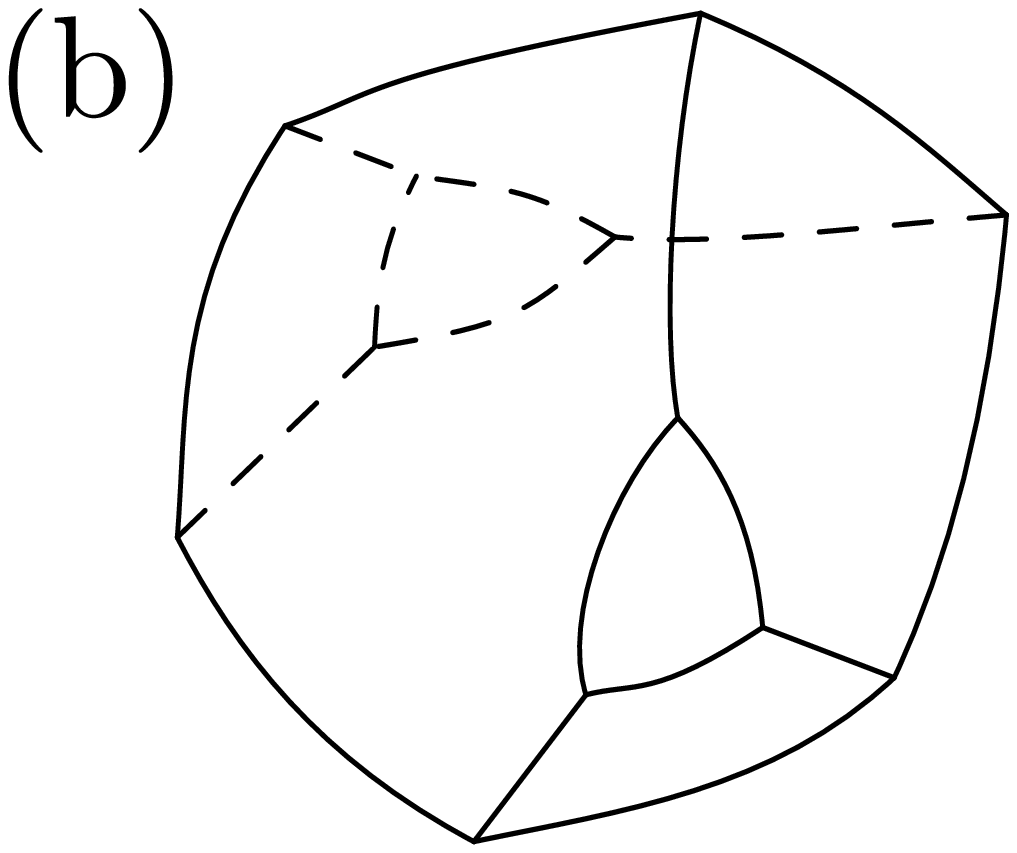}}
\end{tabular}    
\vspace{-5mm}
\end{center}
\caption{Two grains with eight faces but different topologies.}
\label{eightfaces}
\end{figure}

{\it Theory.} The most basic topological property of a grain is its number of neighbors.  In this letter, we use the term {\it neighbors} to refer to pairs of grains that share a common edge or face, in two or three dimensions, respectively.  In two dimensions, the topology of a grain is fully described by its number of edges, which in most cases is equal to its number of neighbors (in exceptional cases, a pair of neighboring grains can share multiple edges).  The arrangement of neighbors in three dimensions, however, is more complicated.  Consider, for example, Fig.~\ref{eightfaces}, which illustrates two grains, each with eight faces.  Although the grains have identical numbers and types of faces, differences in the arrangements of those faces indicate differences in the arrangements of their neighbors.  

We say that two grains have the same {\it topological type}, or topology, if their neighbors can be paired so that neighbors of one grain are themselves neighbors if corresponding neighbors of the other grain are also neighbors.  In two dimensions, each topological type is identified with a natural number.  In three dimensions, each type is identified with a graph isomorphism class \cite{2012lazar,lazar2011evolution}.

{\it Thermodynamics.} A central goal of statistical thermodynamics is understanding the distribution of microstates of a system when only macrostate features are known.  A system of $N$ identical particles confined to a fixed volume in thermal equilibrium with a surrounding fixed-temperature heat bath is a classic example, the canonical ensemble \cite{tuckerman2010statistical}.  What are its possible microstates and what are the probabilities of observing them?  In this example, the probability density $p(\omega)$ of observing the system in microstate $\omega$ depends only on its energy $E_\omega$ and a constant $\beta$, commonly understood as an inverse temperature:
\begin{equation}
p(\omega) = \frac{1}{Z} e^{-\beta E_\omega}.
\label{maineq}
\end{equation}
The partition function $Z = Z(\beta)$ is a normalizing constant which ensures that $p$ is a probability distribution on $\Omega$, the set of all possible microstates.  The exponential dependence of probability on energy results from treating the system and its significantly larger surroundings as an isolated system with fixed energy, and in which all possible microstates are equally probable \cite{aleksandr1949mathematical}.  

These concepts may not initially appear relevant to grain growth for several reasons.  First, unlike in classical statistical mechanics, the energy defined below is not a conserved quantity---the total energy of an isolated system changes with time.  Second, whereas $\beta$ is traditionally interpreted as an inverse temperature, temperature has no obvious physical interpretation in studying the scale-invariant statistical properties of steady-state grain growth microstructures.  We nevertheless suggest that grain growth be considered in this thermodynamic spirit.  In particular, we treat each grain as a separate thermodynamic system whose microstate is described by an energy written solely in terms of its topology.  The probability of a grain having a given topology is then postulated to depend on this energy in a form similar to Eq.~(\ref{maineq}).  We ask that lack of {\it a priori} justification for this approach be momentarily ignored in light of its success in describing the relevant probability distributions. 

{\it Two dimensions.} Although grains in two-dimensional systems are not regular polygons, we consider them so as a first-order approximation.  Adjacent edges of a regular $n$-sided polygon meet at internal angles of $\alpha_n = \pi - 2\pi/n$. Energetic factors in isotropic grain growth, however, cause edges to meet at angles of $\theta_2 = 2\pi/3$.  We therefore define an energy associated with each vertex of an $n$-sided face as the square of the difference between $\alpha_n$ and $\theta_2$, in a manner analogous to a conventional elastic energy.   The total energy associated with an $n$-sided grain is the sum of these energies over its $n$ vertices:
\begin{equation}
E_2(n) = n(\alpha_n - \theta_2)^2.
\label{energy-equation2D}
\end{equation}
Since each angle of a regular hexagon is $\alpha_6 = \theta_2$, the energy associated with the $n=6$ topology is zero. 

Although microstate energies largely determine their probabilities, the manner in which microstates are counted must also be considered.  In particular, if neighbors of a grain are cyclically permuted, or else their order is reversed, then its topology is unchanged, as pairs of grains are neighbors after this transformation only if they were neighbors before it. This identification leads to a corrective factor of $1/s(t_i)$, where $s(t_i)$ is the order of the symmetry group of grain topology $t_i$.  This factor is analogous to the more familiar $1/N!$ factor that arises in systems of $N$ indistinguishable particles, described by Eq.~(\ref{maineq}), which are invariant under the $N!$ permutations belonging to the symmetric group of degree $N$.

In two dimensions, the symmetry group of each regular $n$-gon is the dihedral group with order $2n$, suggesting the following probability distribution of $n$-sided grains:
\begin{equation}
p(n) = \frac{1}{Z}\frac{e^{-\beta E_2(n)}}{2n}, 
\label{fit-eq}
\end{equation}
for some constant $\beta$.  As mentioned before, we are not aware of any physical interpretation of $\beta$, and regard it as a fitting parameter. 

\begin{figure}
\includegraphics[width=0.91\linewidth]{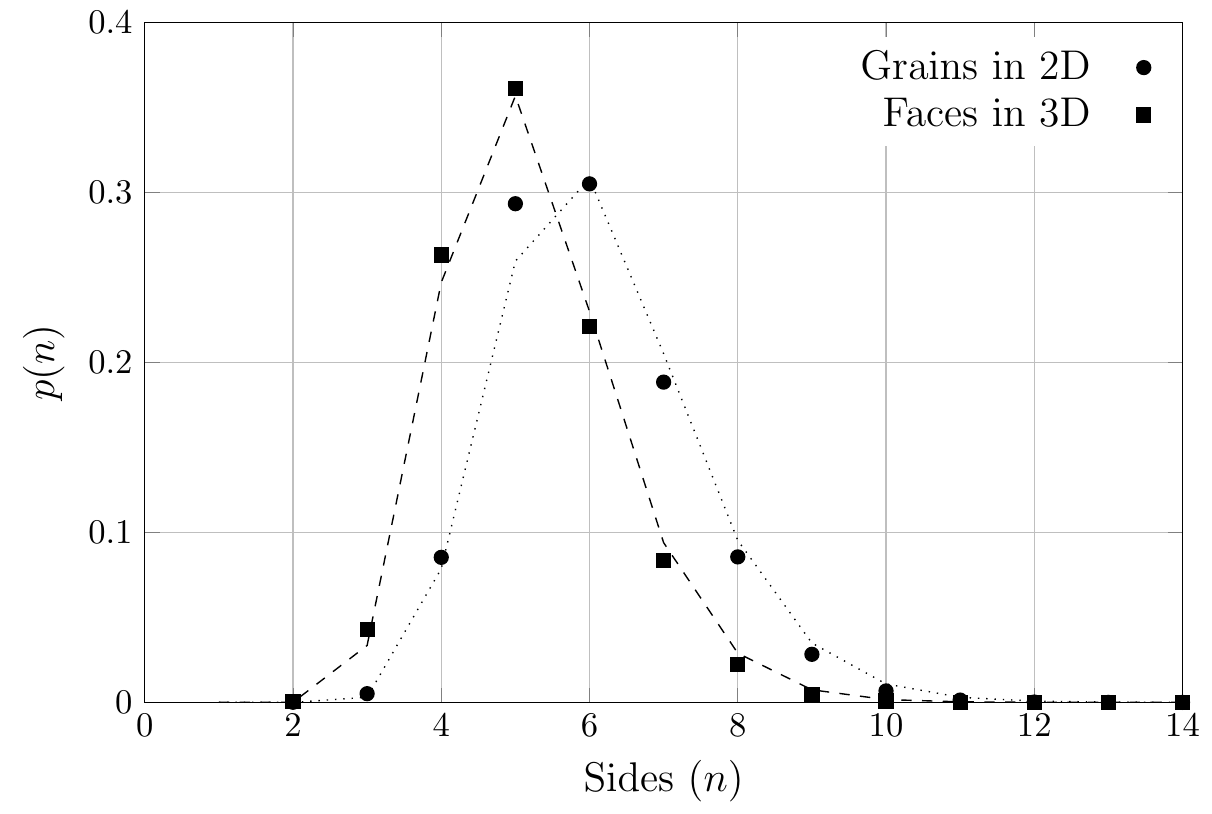}\vspace{-2mm}
\caption{Distribution of $n$-sided grains and faces in two- and three-dimensional steady-state grain growth \cite{mason2015geometric}, respectively, compared with Eqs.~(\ref{fit-eq}) and (\ref{fit-eq3}), with $\beta = 1.62$ and $\beta = 1.29$, for the two systems.  Although these equations are defined only for integer values, we illustrate them as continuous functions to aid visualization; error bars showing standard errors of the mean are smaller than the data points.}
\label{fit-both}
\end{figure}

Figure \ref{fit-both} compares the distribution of grain topologies in steady-state, two-dimensional normal grain-growth microstructures as described by Eq.~(\ref{fit-eq}) with data obtained from prior front-tracking simulations \cite{2010lazar, mason2015geometric}.
A weighted least-squares method finds that the data fit the proposed theory best when $\beta = 1.62$ ($\chi^2 = 0.030$). 
While the equation and the observed values of $p(n)$ do not agree exactly, their similarity in shape suggests that the proposed thermodynamic approach might provide a valuable first-order approximation of the distribution.  

\begin{figure*}
\includegraphics[width=0.47\linewidth]{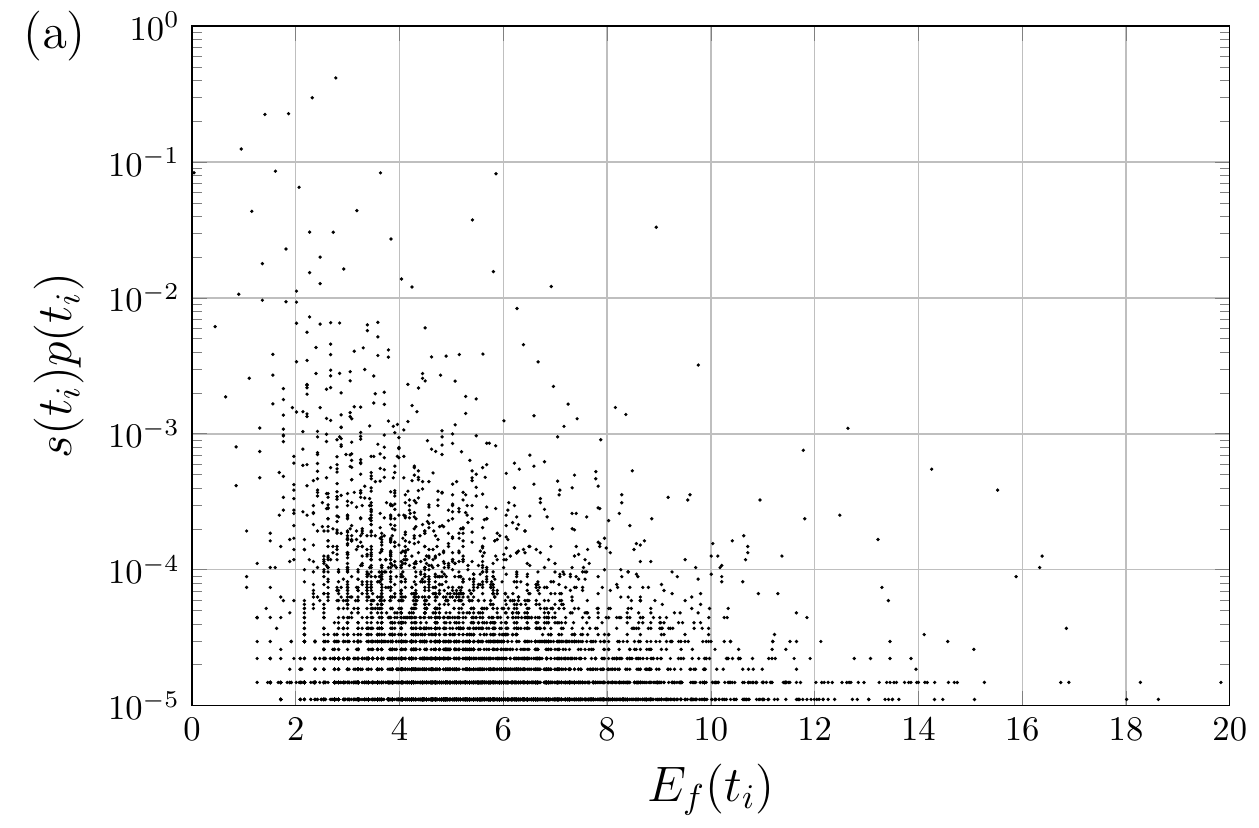} \hspace{3mm}
\includegraphics[width=0.47\linewidth]{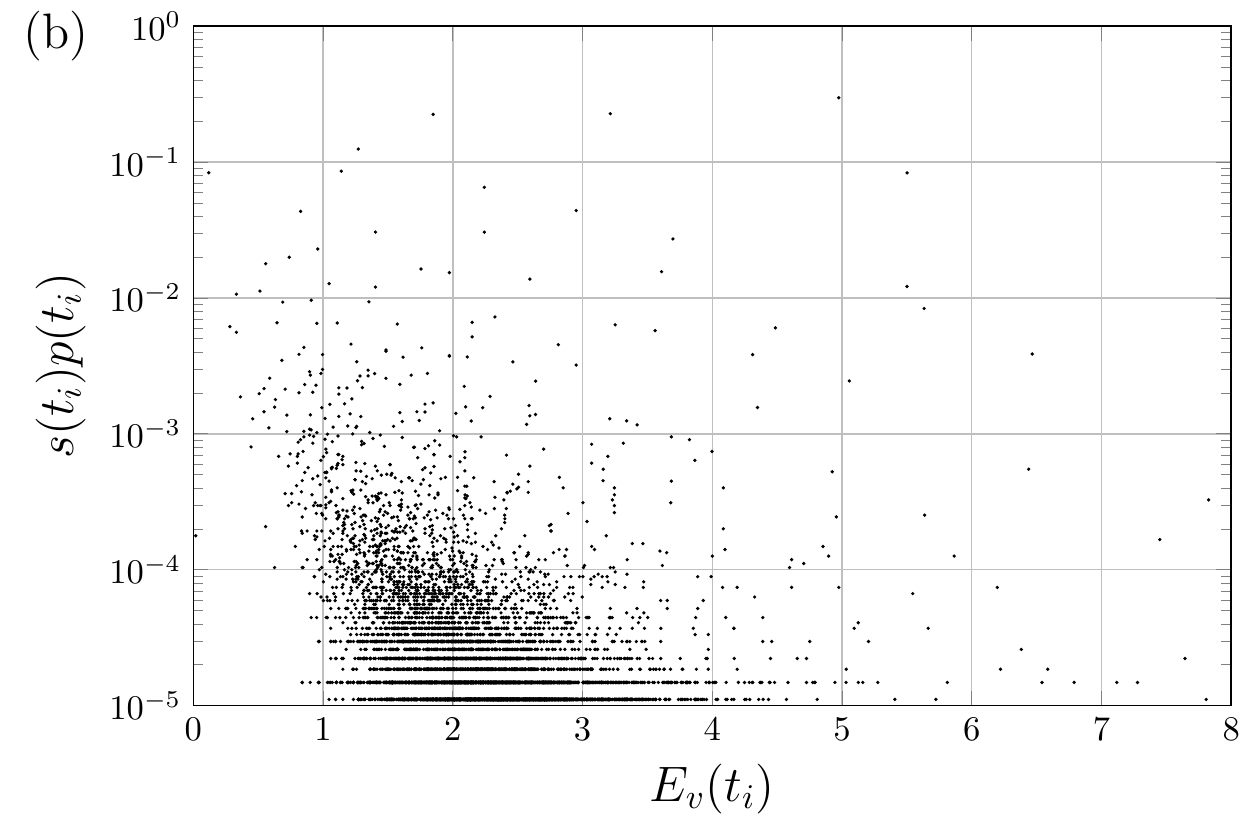} \vspace{-1mm}
\caption{The product of the observed probability $p(t_i)$ and the symmetry group order $s(t_i)$ as a function of energies (a) $E_f$ and (b) $E_v$ for each observed topology $t_i$, as suggested by Eq.~(\ref{pfS}).  Data are taken from three-dimensional front-tracking simulations of steady-state grain growth \cite{2011lazar, mason2015geometric}.}\vspace{-1.5mm}
\label{E_fv_energy}
\end{figure*}

{\it Three dimensions.} Our earlier simulations of three-dimensional grain growth suggested that certain topologies appear more frequently than others, even among those with the same number and types of faces \cite{2012lazar}.  We observed that ``just as curvature flow drives towards geometrically symmetric spheres $\ldots$ it also drives towards topologically symmetric polyhedra.''  We now extend the approach introduced above to analyze three-dimensional systems and to quantify this topological symmetry.  

{\it Distribution of faces.}
We first consider grain faces in three dimensions.  Whereas edges in isotropic, two-dimensional grain growth meet at angles $2\pi/3$, in three dimensions they meet at angles $\theta_3 =  \cos^{-1}(-1/3) \approx 109.5^{\circ}$.  This suggests defining a bending-like energy associated with an $n$-sided face in three dimensions:
\begin{equation}
E_3(n) = n(\alpha_n - \theta_3)^2,
\label{energy-equation23D}
\end{equation}
analogous to the energy defined in Eq.~(\ref{energy-equation2D}); as before, $\alpha_n = \pi - 2\pi/n$.  This energy can be used to estimate the distribution of faces with $n$ sides in three dimensions:
\begin{equation}
p(n) = \frac{1}{Z}\frac{e^{-\beta E_3(n)}}{2n}.
\label{fit-eq3}
\end{equation}

Figure \ref{fit-both} shows steady-state data collected from isotropic grain growth simulations with over 250,000 grains \cite{2011lazar, mason2015geometric}, compared with Eq.~(\ref{fit-eq3}).  A weighted least-squares method finds this equation describes the observed data best when $\beta = 1.29$ ($\chi^2 = 0.009$).  This prediction fits the data more closely than Eq.~(\ref{fit-eq}) did for two-dimensional systems. Unlike in two dimensions, in which $E_2(6) = 0$, in three dimensions, $E_3(n) > 0$ for all $n$, and is minimal when $n=5$.

{\it Distribution of grain topologies.} Topologically-defined energies can also be used to estimate the distribution of topological types in three dimensions.  We define two such energies for each grain topology $t_i$.  The first is a sum of Eq.~(\ref{energy-equation23D}) over all $F$ faces of a grain:
\begin{equation}
E_f(t_i) = \sum_{j=1}^{F} E_3(n_j),
\label{eft}
\end{equation}
where $n_j$ is the number of sides of face $j$.  This energy extends the one defined for polygonal faces to entire grains. The probability of a grain with topology $t_i$ can then be estimated by
\begin{equation}
p(t_i) = \frac{1}{Z}\frac{ e^{-\beta E(t_i)}}{s(t_i)},
\label{pfS}
\end{equation}
where $E(t_i) = E_f(t_i)$, and where $s(t_i)$ is the order of the associated symmetry group; more details about this symmetry group and the algorithm used to calculate its order can be found in Ref.~\cite{1966weinberg2}.  The product $s(t_i) p(t_i)$ is generally reported in the following to emphasize its exponential dependence on energy.

Figure \ref{E_fv_energy}(a) shows the product $s(t_i) p(t_i)$ as a function of $E_f$ for topologies observed in simulations. Those with large $E_f$ appear infrequently, while those with small $E_f$ may appear frequently or infrequently. These data suggest that Eq.~(\ref{pfS}) reasonably approximates the distribution of grain topologies. 

\begin{figure*}
\includegraphics[width=0.98\linewidth]{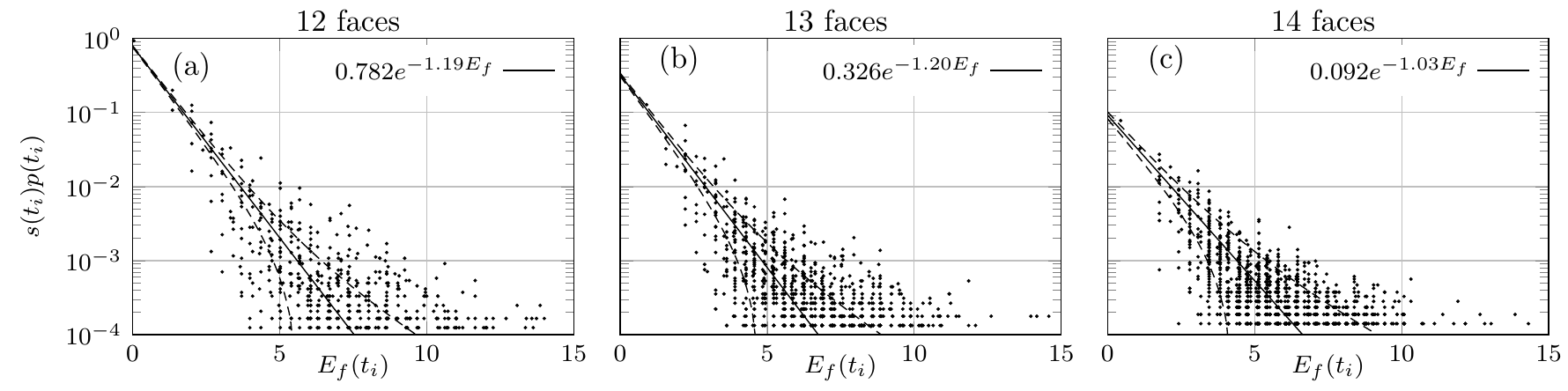}
\includegraphics[width=0.98\linewidth]{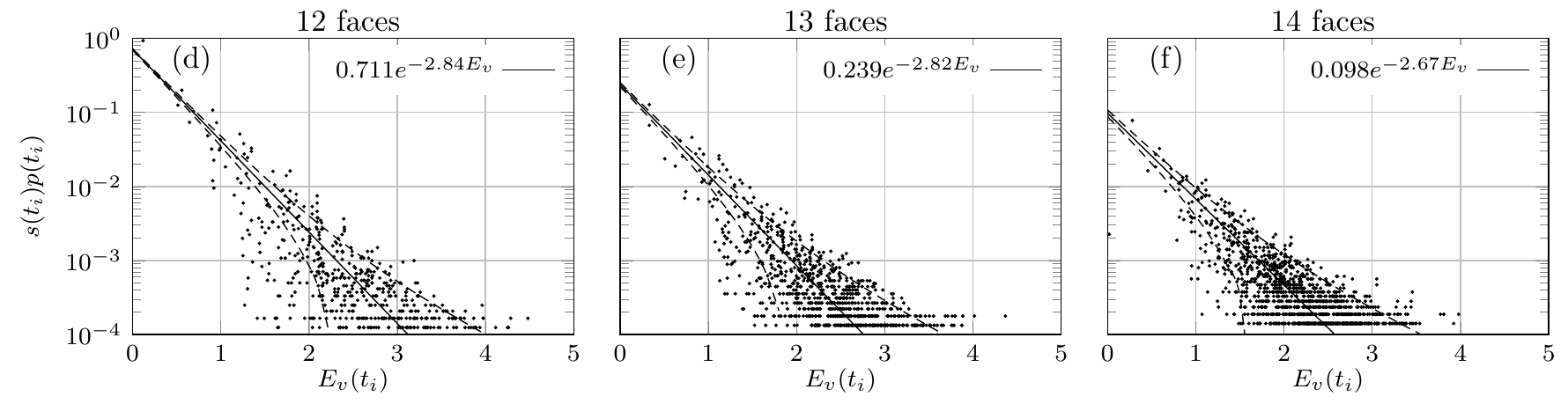}\vspace{-1mm}
\caption{The product of the observed probability $p(t_i)$ and the symmetry group order $s(t_i)$ as a function of (a-c) $E_f$ and of (d-f) $E_v$ for each observed topology $t_i$ with fixed numbers of faces.  Probabilities $p(t_i)$ are normalized so that they sum to 1 for each number of faces. Dashed curves show five standard deviations of the sample mean for the relevant sample size.  Data are taken from simulations of three-dimensional steady-state grain growth \cite{2011lazar, mason2015geometric}.}
\label{EfEv-data}\vspace{-2mm}
\end{figure*}

Although $E_f$ quantifies the energetic favorability of each grain topology, it depends only on the types of faces of a grain, but not on how those faces are arranged.  Such information, however, might yield a more accurate estimate of the distribution of topologies.  For example, the two topologies illustrated in Fig.~\ref{eightfaces} have the same number and types of faces, and hence $E_f$ values, yet the topology illustrated in Fig.~\ref{eightfaces}(a) appears nearly 100 times more frequently than that illustrated in Fig.~\ref{eightfaces}(b).  

We therefore define a second energy to quantify how curvature is distributed over grain vertices.  If three regular $n$-sided polygons meet at a vertex $v$, then the Gaussian curvature concentrated at that vertex is $K_v = 2\pi - (\alpha_{n_1} + \alpha_{n_2} + \alpha_{n_3})$, where $n_j$ is the number of sides of face $j$.  If we approximate each face as a regular polygon, then $K_v$ approximates the actual curvature in purely topological terms. In isotropic grain growth, however, the Gaussian curvature at each vertex is $\hat K = 2\pi - 3 \theta_3$, where $\theta_3 = \cos^{-1}(-1/3)$. We then define the energy at each vertex as the square of the difference between these curvatures, and define the total energy $E_v(t_i)$ of topology $t_i$ as a sum of these energies over its $V$ vertices:
\begin{equation}
E_v(t_i) = \sum_{j=1}^V (K_{v_j} - \hat K)^2
\label{pvS}
\end{equation}
Two grains with the same number and types of faces will generally have identical $E_f$ but different $E_v$.

Figure \ref{E_fv_energy}(b) shows $s(t_i) p(t_i)$ as a function of $E_v$ for each observed grain topology.
Grains with large $E_v$ appear infrequently, while those with low $E_v$ can appear frequently or infrequently. In contrast to Fig.~\ref{E_fv_energy}(a), the predicted probabilities are more scattered.

We next consider the relationship between $E_f$, $E_v$, and $s(t_i)p(t_i)$ when restricted to grains with fixed numbers of faces.  For each fixed number of faces, we use a weighted least-squares method to fit data to a curve of the form $s(t_i)p(t_i) = \frac{1}{Z} e^{-\beta E(t_i)}$.   Figure \ref{EfEv-data} shows data for types with 12, 13, and 14 faces.  These data suggest that $E_f$, $E_v$, and $s(t_i)$ can be used to more accurately estimate the distribution of types when restricted to fixed number of faces.  The only notable outlier appears in Fig.~\ref{EfEv-data}(f) for a point with $E_v \approx 0$, which appears less frequently than predicted.  This point represents the truncated octahedron, which appears only once in the grain-growth simulation dataset.  

\begin{figure*}
\includegraphics[width=0.99\linewidth]{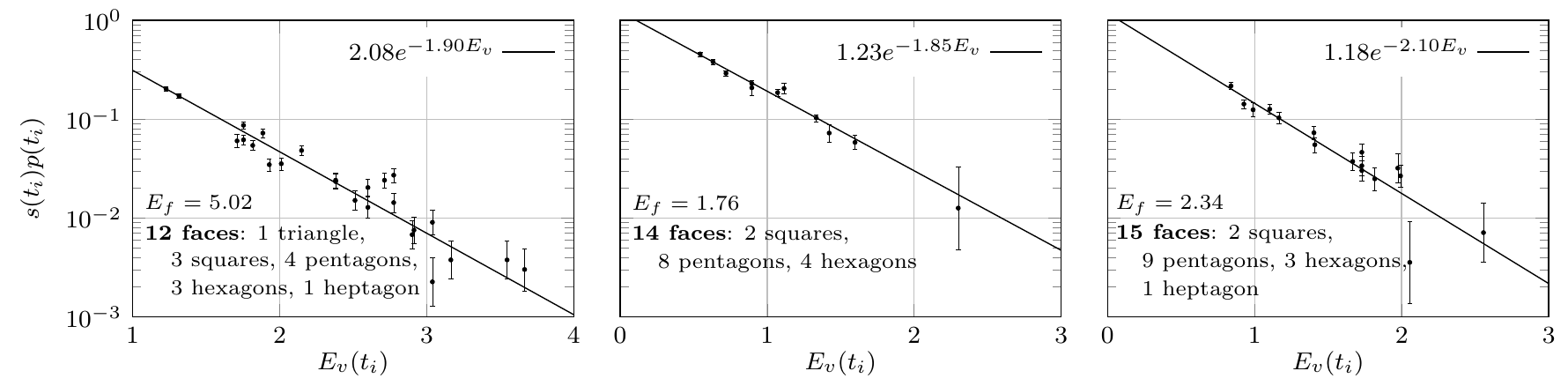}
\caption{Three examples for which a fixed set of faces provides a distribution of topological types.  Probabilities are relative to other samples in the limited dataset; error bars show standard errors of the mean.}\vspace{-3mm}
\label{Efv-data}
\end{figure*}

Finally, we consider sets of grain topologies with identical numbers and types of faces, but in which those faces are arranged differently, thus providing multiple $E_v$ values for fixed $E_f$.  Figure \ref{Efv-data} shows three such datasets, chosen because of their high number of samples of multiple topological types.  In each set, increasing values of $E_v$ are clearly associated with an exponential decrease in $s(t_i)p(t_i)$, suggesting  that $E_v$ and $E_f$ together provide a more accurate prediction of probability than does $E_f$ alone. Specifically, grain topologies in which faces meet in unfavorable ways, as characterized by $E_v$, appear orders of magnitude less frequently than other topologies constructed from identical sets of polygonal faces.

{\it Conclusions.} The most surprising finding of this work is the ability of a topologically-defined ``energy'' to predict the distribution of grain topologies in steady-state, isotropic grain growth.  The similarity between the forms of the energies and distributions in two and three dimensions suggests a common factor governing their behavior.  These energies can be understood as measuring the deviation of realistic grains and their geometries from ideal ones in topological terms.

The relationship between topologically-defined energy, symmetry, and probability is reminiscent of the classical statistical mechanics approach toward analyzing equilibrium systems.  Although grain-growth microstructures are not equilibrium systems, their steady-state properties provide a similar setting for this kind of analysis \cite{fraser1988steady}.  In particular, the existence of an asymptotic state in which scale-invariant properties are statistically constant implies that once dimensional factors are scaled out, microstructure is determined by an energy minimization principle. This is not unusual in systems for which there are large disparities in time scales of different processes; here, the overall coarsening of the microstructure can be considered as ``slow'' while the topological or scale-free microstructural evolution is ``fast''. Hence, late-time evolution of grain growth can be described using a microstructural Born-Oppenheimer approximation. 

While the energies suggested here can be thought of as approximating bending energies, other topologically-defined energies might also be considered.  For example, a twisting energy can be defined along grain edges to quantify the strain resulting from differences in face arrangements at alternate ends. Further, while the context of the current study is grain growth in polycrystalline metals, the suggested approach may find application in understanding data collected in studies of polyhedra-shaped cells in other systems, such as bubbles in soap foams \cite{matzke1946three} and polyhedrocytes in blood clots and thrombi \cite{tutwiler2018shape}. 

Finally, the approach introduced in this letter might be compared with that recently proposed by Lutz {\it et al.}~\cite{lutz2017roundness}.  In both, an energy is defined in purely topological terms to capture the favorability of each topological type, and is then used to estimate its probability.  One strength of the approach suggested here is its connection to classical statistical mechanics and its exponential relationship between energy, symmetry, and probability.  

{\it Acknowledgments.} E.A.L. and D.J.S. acknowledge the generous support of the U.S. National Science Foundation, through Award DMR-1507013.  The research contribution of D. J. S. was also sponsored, in part, by the Army Research Office and was accomplished under Grant Number W911NF-19-1-0263. The views and conclusions contained in this document are those of the authors and should not be interpreted as representing the official policies, either expressed or implied, of the Army Research Office or the U.S. Government. The U.S. Government is authorized to reproduce and distribute reprints for Government purposes notwithstanding any copyright notation herein.

\bibliographystyle{ieeetr} 
\bibliography{refs}       

\end{document}